\documentclass[aps,prd,preprint,tightenlines,superscriptaddress,showpacs,byrevtex,floatfix,nofootinbib]{revtex4}
\usepackage{graphicx} 
\usepackage{dcolumn}  
\usepackage{bm}
\usepackage{amsmath}
\usepackage{color}
\usepackage{multirow}

\graphicspath{{ps}}

\newcommand{\hphm}{\ensuremath{B^{0} \to h^{+}h^{-}}}
\newcommand{\pippim}{\ensuremath{B^{0} \to \pi^{+}\pi^{-}}}
\newcommand{\kppim}{\ensuremath{B^{0} \to K^{+}\pi^{-}}}
\newcommand{\kmpip}{\ensuremath{\bar{B}^{0} \to K^{-}\pi^{+}}}
\newcommand{\kpi}{\ensuremath{B^{0} \to K^{\pm}\pi^{\mp}}}
\newcommand{\kpkm}{\ensuremath{B^{0} \to K^{+}K^{-}}}
\newcommand{\dpi}{\ensuremath{B^{+} \to \bar{D}^{0} [K^{+}\pi^{-}] \pi^{+}}}

\newcommand{\epem}{\ensuremath{e^{+} e^{-}}}
\newcommand{\qqbar}{\ensuremath{q \bar{q}}}
\newcommand{\BBbar}{\ensuremath{B \bar{B}}}
\newcommand{\BzBzb}{\ensuremath{B^{0} \bar{B}^{0}}}
\newcommand{\BpBm}{\ensuremath{B^{+} B^{-}}}

\newcommand{\pip}{\ensuremath{\pi^{+}}}
\newcommand{\pim}{\ensuremath{\pi^{-}}}
\newcommand{\pimp}{\ensuremath{\pi^{\mp}}}
\newcommand{\piz}{\ensuremath{\pi^{0}}}
\newcommand{\aone}{\ensuremath{a_{1}^{\pm}}}

\newcommand{\Kp}{\ensuremath{K^{+}}}
\newcommand{\Km}{\ensuremath{K^{-}}}
\newcommand{\Kpm}{\ensuremath{K^{\pm}}}

\newcommand{\Kz}{\ensuremath{K^{0}}}
\newcommand{\Bp}{\ensuremath{B^{+}}}
\newcommand{\Bz}{\ensuremath{B^{0}}}
\newcommand{\Bzb}{\ensuremath{\bar{B}^{0}}}
\newcommand{\Ups}{\ensuremath{\Upsilon(4S)}}

\newcommand{\Brec}{\ensuremath{B^{0}_{\rm Rec}}}
\newcommand{\Btag}{\ensuremath{B^{0}_{\rm Tag}}}

\newcommand{\Mbc}{\ensuremath{M_{\rm bc}}}
\newcommand{\De}{\ensuremath{\Delta E}}
\newcommand{\Fsb}{\ensuremath{{\cal F}_{b \bar b/q \bar q}}}
\newcommand{\Lp}{\ensuremath{{\cal L}^{+}_{K/\pi}}}
\newcommand{\Lm}{\ensuremath{{\cal L}^{-}_{K/\pi}}}
\newcommand{\Lpm}{\ensuremath{{\cal L}^{\pm}_{K/\pi}}}
\newcommand{\Dt}{\ensuremath{\Delta t}}
\newcommand{\Dz}{\ensuremath{\Delta z}}

\newcommand{\taub}{\ensuremath{\tau_{B^{0}}}}
\newcommand{\Dw}{\ensuremath{\Delta w}}
\newcommand{\Dmd}{\ensuremath{\Delta m_{d}}}
\newcommand{\Acp}{\ensuremath{{\cal A}_{CP}}}
\newcommand{\Scp}{\ensuremath{{\cal S}_{CP}}}
\newcommand{\phitwo}{\ensuremath{\phi_{2}}}
\newcommand{\phitwoeff}{\ensuremath{\phi^{\rm eff}_{2}}}

\begin{document}

\vspace*{-3\baselineskip}

\begin{minipage}[]{0.6\columnwidth}
  \includegraphics[height=3.0cm,width=!]{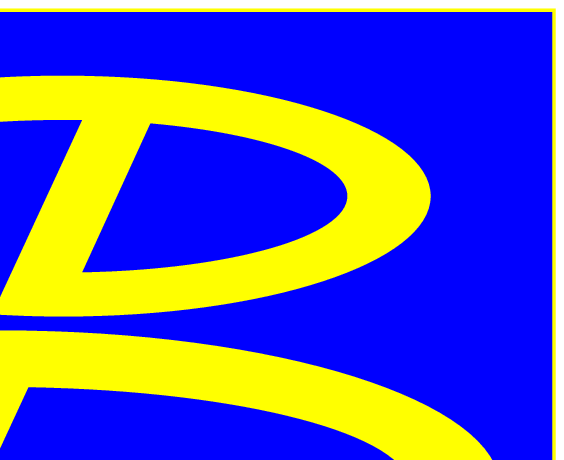}
\end{minipage}
\begin{minipage}[]{0.4\columnwidth}
  \hbox{Belle Preprint 2013-11}
  \hbox{KEK Preprint 2013-21}
\end{minipage}


\title{ \quad\\[1.0cm] Measurement of the $\bm{CP}$ violation parameters in\\ $\bm{\pippim}$ decays}

\noaffiliation
\affiliation{University of the Basque Country UPV/EHU, 48080 Bilbao}
\affiliation{University of Bonn, 53115 Bonn}
\affiliation{Budker Institute of Nuclear Physics SB RAS and Novosibirsk State University, Novosibirsk 630090}
\affiliation{Faculty of Mathematics and Physics, Charles University, 121 16 Prague}
\affiliation{Chiba University, Chiba 263-8522}
\affiliation{University of Cincinnati, Cincinnati, Ohio 45221}
\affiliation{Deutsches Elektronen--Synchrotron, 22607 Hamburg}
\affiliation{Justus-Liebig-Universit\"at Gie\ss{}en, 35392 Gie\ss{}en}
\affiliation{Gifu University, Gifu 501-1193}
\affiliation{II. Physikalisches Institut, Georg-August-Universit\"at G\"ottingen, 37073 G\"ottingen}
\affiliation{Gyeongsang National University, Chinju 660-701}
\affiliation{Hanyang University, Seoul 133-791}
\affiliation{University of Hawaii, Honolulu, Hawaii 96822}
\affiliation{High Energy Accelerator Research Organization (KEK), Tsukuba 305-0801}
\affiliation{Hiroshima Institute of Technology, Hiroshima 731-5193}
\affiliation{Ikerbasque, 48011 Bilbao}
\affiliation{Indian Institute of Technology Guwahati, Assam 781039}
\affiliation{Indian Institute of Technology Madras, Chennai 600036}
\affiliation{Institute of High Energy Physics, Chinese Academy of Sciences, Beijing 100049}
\affiliation{Institute of High Energy Physics, Vienna 1050}
\affiliation{Institute for High Energy Physics, Protvino 142281}
\affiliation{INFN - Sezione di Torino, 10125 Torino}
\affiliation{Institute for Theoretical and Experimental Physics, Moscow 117218}
\affiliation{J. Stefan Institute, 1000 Ljubljana}
\affiliation{Kanagawa University, Yokohama 221-8686}
\affiliation{Institut f\"ur Experimentelle Kernphysik, Karlsruher Institut f\"ur Technologie, 76131 Karlsruhe}
\affiliation{Kavli Institute for the Physics and Mathematics of the Universe (WPI), University of Tokyo, Kashiwa 277-8583}
\affiliation{Korea Institute of Science and Technology Information, Daejeon 305-806}
\affiliation{Korea University, Seoul 136-713}
\affiliation{Kyungpook National University, Daegu 702-701}
\affiliation{\'Ecole Polytechnique F\'ed\'erale de Lausanne (EPFL), Lausanne 1015}
\affiliation{Faculty of Mathematics and Physics, University of Ljubljana, 1000 Ljubljana}
\affiliation{Luther College, Decorah, Iowa 52101}
\affiliation{University of Maribor, 2000 Maribor}
\affiliation{Max-Planck-Institut f\"ur Physik, 80805 M\"unchen}
\affiliation{School of Physics, University of Melbourne, Victoria 3010}
\affiliation{Moscow Physical Engineering Institute, Moscow 115409}
\affiliation{Moscow Institute of Physics and Technology, Moscow Region 141700}
\affiliation{Graduate School of Science, Nagoya University, Nagoya 464-8602}
\affiliation{Kobayashi-Maskawa Institute, Nagoya University, Nagoya 464-8602}
\affiliation{Nara Women's University, Nara 630-8506}
\affiliation{National Central University, Chung-li 32054}
\affiliation{National United University, Miao Li 36003}
\affiliation{Department of Physics, National Taiwan University, Taipei 10617}
\affiliation{H. Niewodniczanski Institute of Nuclear Physics, Krakow 31-342}
\affiliation{Nippon Dental University, Niigata 951-8580}
\affiliation{Niigata University, Niigata 950-2181}
\affiliation{University of Nova Gorica, 5000 Nova Gorica}
\affiliation{Osaka City University, Osaka 558-8585}
\affiliation{Pacific Northwest National Laboratory, Richland, Washington 99352}
\affiliation{Panjab University, Chandigarh 160014}
\affiliation{University of Pittsburgh, Pittsburgh, Pennsylvania 15260}
\affiliation{Research Center for Electron Photon Science, Tohoku University, Sendai 980-8578}
\affiliation{University of Science and Technology of China, Hefei 230026}
\affiliation{Seoul National University, Seoul 151-742}
\affiliation{Soongsil University, Seoul 156-743}
\affiliation{Sungkyunkwan University, Suwon 440-746}
\affiliation{School of Physics, University of Sydney, NSW 2006}
\affiliation{Tata Institute of Fundamental Research, Mumbai 400005}
\affiliation{Excellence Cluster Universe, Technische Universit\"at M\"unchen, 85748 Garching}
\affiliation{Toho University, Funabashi 274-8510}
\affiliation{Tohoku Gakuin University, Tagajo 985-8537}
\affiliation{Tohoku University, Sendai 980-8578}
\affiliation{Department of Physics, University of Tokyo, Tokyo 113-0033}
\affiliation{Tokyo Institute of Technology, Tokyo 152-8550}
\affiliation{Tokyo Metropolitan University, Tokyo 192-0397}
\affiliation{Tokyo University of Agriculture and Technology, Tokyo 184-8588}
\affiliation{University of Torino, 10124 Torino}
\affiliation{CNP, Virginia Polytechnic Institute and State University, Blacksburg, Virginia 24061}
\affiliation{Wayne State University, Detroit, Michigan 48202}
\affiliation{Yamagata University, Yamagata 990-8560}
\affiliation{Yonsei University, Seoul 120-749}
  \author{J.~Dalseno}\affiliation{Max-Planck-Institut f\"ur Physik, 80805 M\"unchen}\affiliation{Excellence Cluster Universe, Technische Universit\"at M\"unchen, 85748 Garching} 
  \author{K.~Prothmann}\affiliation{Max-Planck-Institut f\"ur Physik, 80805 M\"unchen}\affiliation{Excellence Cluster Universe, Technische Universit\"at M\"unchen, 85748 Garching} 
  \author{C.~Kiesling}\affiliation{Max-Planck-Institut f\"ur Physik, 80805 M\"unchen} 
  \author{I.~Adachi}\affiliation{High Energy Accelerator Research Organization (KEK), Tsukuba 305-0801} 
  \author{H.~Aihara}\affiliation{Department of Physics, University of Tokyo, Tokyo 113-0033} 
  \author{K.~Arinstein}\affiliation{Budker Institute of Nuclear Physics SB RAS and Novosibirsk State University, Novosibirsk 630090} 
  \author{D.~M.~Asner}\affiliation{Pacific Northwest National Laboratory, Richland, Washington 99352} 
  \author{V.~Aulchenko}\affiliation{Budker Institute of Nuclear Physics SB RAS and Novosibirsk State University, Novosibirsk 630090} 
  \author{T.~Aushev}\affiliation{Institute for Theoretical and Experimental Physics, Moscow 117218} 
  \author{A.~M.~Bakich}\affiliation{School of Physics, University of Sydney, NSW 2006} 
% ...
\collaboration{The Belle Collaboration}

\begin{abstract}
  We present a measurement of the charge-parity $(CP)$ violating parameters in \pippim\ decays. The results are obtained from the final data sample containing $772 \times 10^{6}$ \BBbar\ pairs collected at the \Ups\ resonance with the Belle detector at the KEKB asymmetric-energy \epem\ collider. We obtain the $CP$ violation parameters
  $$
  \begin{array}{rcl}
    \Acp(\pippim) \!\!&=&\!\! +0.33 \pm 0.06 \textrm{ (stat)} \pm 0.03 \textrm{ (syst)},\\
    \Scp(\pippim) \!\!&=&\!\! -0.64 \pm 0.08 \textrm{ (stat)} \pm 0.03 \textrm{ (syst)},\\
  \end{array}
  $$
  where \Acp\ and \Scp\ represent the direct and mixing-induced $CP$ asymmetries in \pippim\ decays, respectively. Using an isospin analysis including results from other Belle measurements, we find $23.8^{\circ} < \phitwo < 66.8^{\circ}$ is disfavored at the $1\sigma$ level, where \phitwo\ is one of the three interior angles of the Cabibbo-Kobayashi-Maskawa unitarity triangle related to $B_{u,d}$ decays.

\end{abstract}


\maketitle

\tighten

{\renewcommand{\thefootnote}{\fnsymbol{footnote}}}
\setcounter{footnote}{0}

\section{Introduction}
Violation of the combined charge-parity symmetry ($CP$ violation) in the standard model (SM) arises from a single irreducible phase in the Cabibbo-Kobayashi-Maskawa (CKM) quark-mixing matrix~\cite{Cabibbo,KM}. A main objective of the Belle experiment at KEK, Japan, is to over-constrain the unitarity triangle of the CKM matrix related to $B_{u,d}$ decays. This permits a precision test of the CKM mechanism for $CP$ violation as well as the search for new physics (NP) effects. Mixing-induced $CP$ violation in the $B$ sector has been clearly established by Belle~\cite{jpsiks_Belle1,jpsiks_Belle2} and BaBar~\cite{jpsiks_BABAR1,jpsiks_BABAR2} in the $\bar b \rightarrow \bar c c \bar s$ induced decay $\Bz \rightarrow J/\psi \Kz$. There are many other modes that may provide additional information on various $CP$ violating parameters.
\begin{figure}
  \centering
  \includegraphics[height=120pt,width=!]{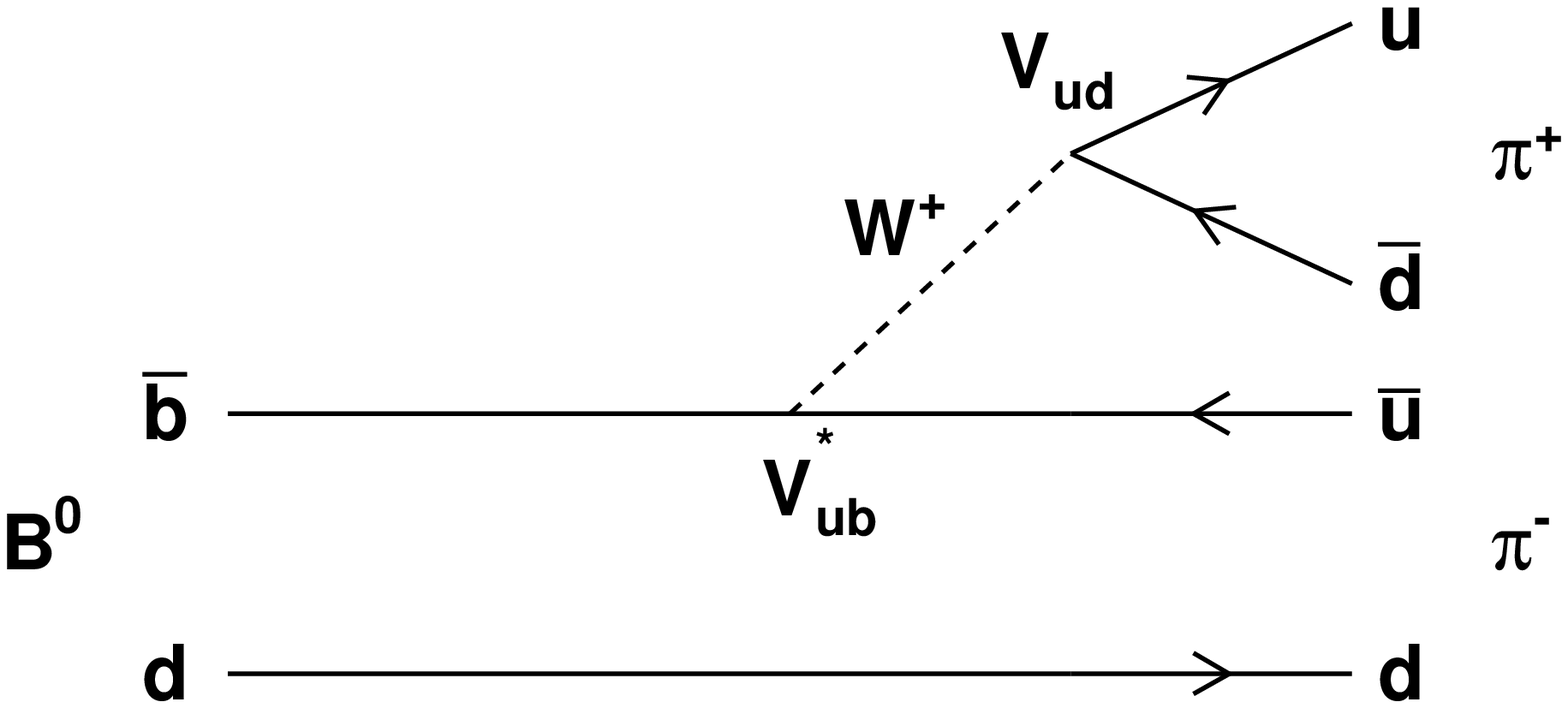}
  \includegraphics[height=120pt,width=!]{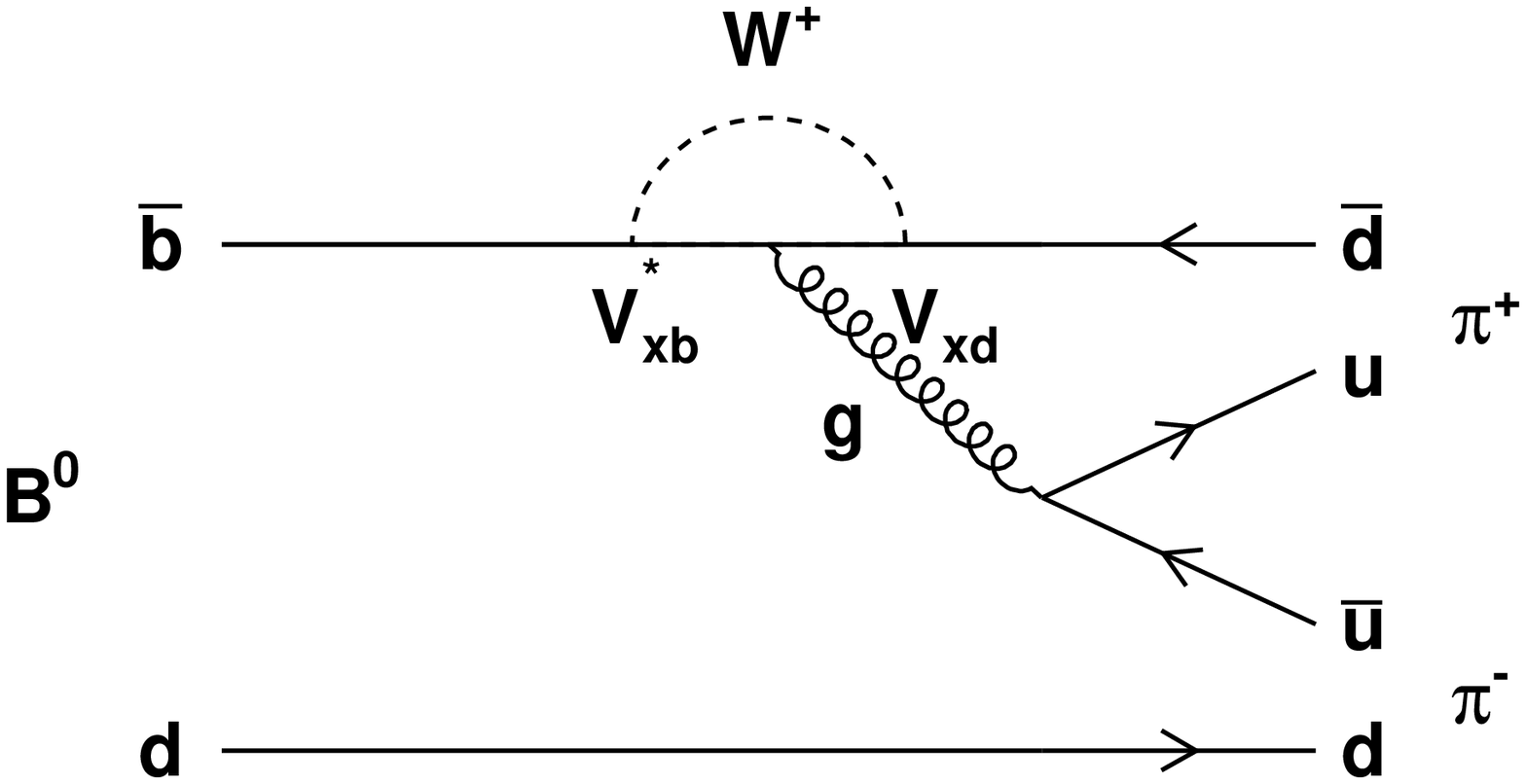}
  \put(-444,110){(a)}
  \put(-218,110){(b)}
  \caption{Leading-order Feynman diagrams for \pippim\ decays. (a) depicts the dominant first-order amplitude (tree) while (b) shows the second-order loop (penguin) diagram. In the penguin diagram, the subscript $x$ in $V_{xb}$ refers to the flavor of the intermediate-state quark $(x=u,c,t)$.}
  \label{fig_hh}
\end{figure}

Decays that proceed predominantly through the $\bar b \rightarrow \bar u u \bar d$ transition are sensitive to the interior angle of the unitarity triangle $\phitwo \equiv \arg(-V_{td}V^{*}_{tb})/(V_{ud}V^{*}_{ub})$\footnote{Another notation, $\alpha \equiv \arg(-V_{td}V^{*}_{tb})/(V_{ud}V^{*}_{ub})$, also exists in literature.}. This paper describes a measurement of $CP$ violation parameters in \pippim\ decays, whose dominant amplitudes are shown in Fig.~\ref{fig_hh}. Belle, BaBar and LHCb have reported time-dependent $CP$ asymmetries in related modes including $\Bz \rightarrow \pip \pim$~\cite{pipi_Belle,pipi_BaBar,pipi_LHCb}, $(\rho \pi)^{0}$~\cite{rhopi_Belle,rhopi_BaBar}, $\rho^{+} \rho^{-}$~\cite{rhorho_Belle,rhorho_BaBar} and $\aone \pi^{\mp}$~\cite{a1pi_Belle,a1pi_BaBar}.

The decay of the \Ups\ can produce a \BzBzb\ pair in a coherent quantum-mechanical state, from which one meson (\Brec) may be reconstructed in the $\pip \pim$ decay mode. This decay mode does not determine whether the \Brec\ decayed as a \Bz\ or as a \Bzb. The $b$ flavor of the other $B$ meson (\Btag), however, can be identified using information from the remaining charged particles and photons. This dictates the flavor of \Brec\ as it must be opposite that of the \Btag\ flavor at the time \Btag\ decays. The proper time interval between \Brec\ and \Btag, which decay at time $t_{\rm Rec}$ and $t_{\rm Tag}$, respectively, is defined as $\Dt \equiv t_{\rm Rec} - t_{\rm Tag}$ measured in the \Ups\ frame. For the case of coherent \BzBzb\ pairs, the time-dependent decay rate for a $CP$ eigenstate when \Btag\ possesses flavor $q$, where \Bz\ has $q=+1$ and \Bzb\ has $q=-1$, is given by
\begin{equation}
  {\cal P}(\Dt, q) = \frac{e^{-|\Dt|/\taub}}{4\taub} \biggl\{1 + q \biggl[\Acp \cos \Dmd \Dt + \Scp\sin \Dmd \Dt\biggr]\biggr\}.
\end{equation}
Here, \taub\ is the \Bz\ lifetime and \Dmd\ is the mass difference between the two mass eigenstates of the neutral $B$ meson. This time dependence assumes $CPT$ invariance, no $CP$ violation in the mixing, and that the difference in decay rates between the two mass eigenstates is negligible. The parameter \Acp\ measures the direct $CP$ violation, while \Scp\ is a measure of the amount of mixing-induced $CP$ violation.

 In the limit that only the dominant tree amplitude contributes, no flavor-dependent direct $CP$ violation is expected and \Scp\ is $\sin2\phitwo$. However, in the \pippim\ final state and other $\bar b \rightarrow \bar u u \bar d$ self-conjugate modes, the value of \phitwo\ is shifted by an amount $\Delta \phitwo$, due to the presence of additional penguin contributions that interfere with the dominant tree contribution (see Fig.~\ref{fig_hh}). Thus, the observable mixing-induced $CP$ parameter becomes $\Scp = \sqrt{1 - \Acp^2}\sin (2 \phitwo + 2 \Delta \phitwo)$.

Despite penguin contamination, it is still possible to determine $\phitwo$ in $\Bz \rightarrow \pip \pim$ with an $SU(2)$ isospin analysis~\cite{theory_su2} by considering the set of $B \rightarrow \pi\pi$ decays into the three possible charge states for the pions. Here, the two pions in $\Bp \rightarrow \pip \piz$ decays must have a total isospin of $I=1$ or $I=2$, since $I_{3} = 1$. For the penguin contributions, only $I=0$ or $I=1$ is possible because the gluon is an isospin singlet carrying $I=0$. However, $I=1$ is forbidden by Bose-Einstein statistics; thus, strong loop decays cannot contribute and hence $\Bp \rightarrow \pip \piz$ decays only through the tree diagram in the limit of negligible electroweak penguins.

\begin{figure}
  \centering
  \includegraphics[height=100pt,width=!]{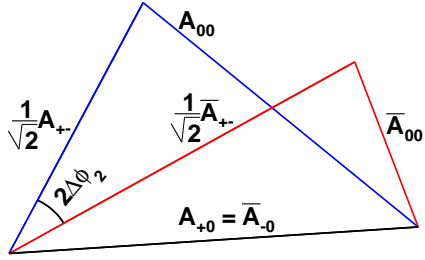}
  \caption{Complex isospin triangles from which $\Delta \phitwo$ can be determined.}
  \label{fig_iso}
\end{figure}

The complex $\Bz \rightarrow \pi\pi$ and $\Bzb \rightarrow \pi\pi$ decay amplitudes obey the relations
\begin{equation}
  A_{+0} = \frac{1}{\sqrt{2}}A_{+-} + A_{00}, \;\;\;\; \bar{A}_{-0} = \frac{1}{\sqrt{2}}\bar{A}_{+-} + \bar{A}_{00},
  \label{eq_iso}
\end{equation}
respectively, where the subscripts refer to the combination of the pion charges. The decay amplitudes can be represented as the triangles shown in Fig.~\ref{fig_iso}. As $\Bp \rightarrow \pip \piz$ is a pure tree mode, these triangles share the same base, $A_{+0}=\bar{A}_{-0}$, and $\Delta \phitwo$ can be determined from the difference between the two triangles. These triangles and \phitwo\ can be fully determined from the branching fractions, ${\cal B}(\Bz \rightarrow \pip \pim)$, ${\cal B}(\Bz \rightarrow \piz\piz)$ and ${\cal B}(\Bp \rightarrow \pip \piz)$, and the $CP$ violation parameters, $\Acp(\Bz \rightarrow \pip \pim)$, $\Scp(\Bz \rightarrow \pip\pim)$ and $\Acp(\Bz \rightarrow \piz\piz)$. This method has an eightfold discrete ambiguity in the determination of \phitwo, which arises from the four triangle orientations about $A_{+0}$ and the two solutions of \phitwoeff\ in the measurement of \Scp.

Belle, BaBar and LHCb have reported measurements~\cite{pipi_Belle,pipi_BaBar,pipi_LHCb}, summarized in Table~\ref{tab_hh_prev}, of the $CP$ violation parameters reported here. The previous Belle measurements were based on a sample of 535 million \BBbar\ pairs and are superseded by the analysis presented here.
\begin{table}
  \centering
  \caption{Summary of $CP$ violation parameters obtained by Belle~\cite{pipi_Belle}, BaBar~\cite{pipi_BaBar} and LHCb~\cite{pipi_LHCb}. For all parameters, the first uncertainty is statistical and the second is systematic. The Belle value for \Acp\ is marginally consistent ($1.9\sigma$) with the BaBar and LHCb measurements.}
  \begin{tabular}
    {@{\hspace{0.5cm}}c@{\hspace{0.25cm}}  @{\hspace{0.25cm}}c@{\hspace{0.5cm}}  @{\hspace{0.25cm}}c@{\hspace{0.5cm}}  @{\hspace{0.25cm}}c@{\hspace{0.5cm}}  @{\hspace{0.25cm}}c@{\hspace{0.5cm}}}
    \hline \hline
    Parameter & Belle & BaBar & LHCb\\
    & ($535\times10^6$ \BBbar\ pairs) & ($467\times10^6$ \BBbar\ pairs) & (0.7 fb$^{-1}$)\\
    \hline
    $\Acp(\pippim)$ & $+0.55 \pm 0.08 \pm 0.05$ & $+0.25 \pm 0.08 \pm 0.02$ & $+0.11 \pm 0.21 \pm 0.03$\\
    $\Scp(\pippim)$ & $-0.61 \pm 0.10 \pm 0.04$ & $-0.68 \pm 0.10 \pm 0.03$ & $-0.56 \pm 0.17 \pm 0.03$\\
    \hline \hline
  \end{tabular}
  \label{tab_hh_prev}
\end{table}

In Sec.~\ref{Data Set And Belle Detector}, we briefly describe the data set and Belle detector. We explain the selection criteria used to identify signal candidates and suppress backgrounds in Sec.~\ref{Event Selection}, followed by the fit method used to extract the signal component in Sec.~\ref{Event Model}. In Sec.~\ref{Fit Result}, the results of the fit are presented along with a discussion of the systematic uncertainties in Sec.~\ref{Systematic Uncertainties}. Finally, our conclusions are given in Sec.~\ref{Conclusion}.

\section{Data Set And Belle Detector}
\label{Data Set And Belle Detector}
This measurement of the $CP$ violation parameters in \pippim\ decays is based on the final data sample containing $772 \times 10^{6}$ \BBbar\ pairs collected with the Belle detector at the KEKB asymmetric-energy \epem\ ($3.5$ on $8~{\rm GeV}$) collider~\cite{KEKB}. At the \Ups\ resonance ($\sqrt{s}=10.58$~GeV), the Lorentz boost of the produced \BBbar\ pairs is $\beta\gamma =0.425$ nearly along the $+z$ direction, which is opposite the positron beam direction. We also use a $100 \; {\rm fb}^{-1}$ data sample recorded at 60 MeV below the \Ups\ resonance, referred to as off-resonance data, for continuum ($e^+ e^- \to q\bar{q}$, where $q = d,\ u,\ s,\ c$) background studies.

The Belle detector is a large-solid-angle magnetic
spectrometer that consists of a silicon vertex detector (SVD),
a 50-layer central drift chamber (CDC), an array of
aerogel threshold Cherenkov counters (ACC),  
a barrel-like arrangement of time-of-flight
scintillation counters (TOF), and an electromagnetic calorimeter (ECL)
comprising CsI(Tl) crystals located inside 
a superconducting solenoid coil that provides a 1.5~T
magnetic field.  An iron flux-return located outside of
the coil is instrumented to detect $K_L^0$ mesons and to identify
muons (KLM).  The detector
is described in detail elsewhere~\cite{Belle}.
Two inner detector configurations were used. A 2.0-cm-radius beampipe
and a three-layer silicon vertex detector (SVD1) were used for the first sample
of $152 \times 10^6 B\bar{B}$ pairs, while a 1.5-cm-radius beampipe, a four-layer
silicon detector (SVD2) and a small-cell inner drift chamber were used to record  
the remaining $620 \times 10^6 B\bar{B}$ pairs~\cite{svd2}. We use a GEANT-based
 Monte Carlo (MC) simulation to model the response of the detector and to determine
 its acceptance~\cite{GEANT}.

\section{Event Selection}
\label{Event Selection}
The decay channel \pippim\ is reconstructed from two oppositely charged tracks. Charged tracks are identified using a loose requirement on the distance of closest approach with respect to the interaction point (IP) along the beam direction, $|dz| < 4.0 \; {\rm cm}$, and in the transverse plane, $dr < 0.4 \; {\rm cm}$. Additional SVD requirements of at least two $z$ hits and one $r-\phi$ hit~\cite{ResFunc} are imposed on all charged tracks so that a good quality vertex of the reconstructed $B$ candidate can be determined. Using information obtained from the CDC, ACC and TOF, particle identification (PID) is determined from a likelihood ratio ${\cal L}_{i/j} \equiv {\cal L}_{i}/({\cal L}_{i} + {\cal L}_{j})$. Here, ${\cal L}_{i}$ (${\cal L}_{j}$) is the likelihood that the particle is of type $i$ ($j$). To suppress background due to electron misidentification, ECL information is used to veto particles consistent with the electron hypothesis. The PID ratios of the two charged tracks \Lpm, are used in the fit model to discriminate among the three possible two-body channels: $\Bz \to \pip \pim$, $\Bz \to \Kp \pim$ and $\Bz \to \Kp \Km$.

Reconstructed $B$ candidates are identified with two nearly uncorrelated kinematic variables: the beam-energy-constrained mass $\Mbc \equiv \sqrt{(E^{\rm CMS}_{\rm beam})^{2} - (p^{\rm CMS}_{B})^{2}}$ and the energy difference $\De \equiv E^{\rm CMS}_{B} - E^{\rm CMS}_{\rm beam}$, where $E^{\rm CMS}_{\rm beam}$ is the beam energy and $E^{\rm CMS}_{B}$ ($p^{\rm CMS}_{B}$) is the energy (momentum) of the $B$ meson, all evaluated in the $e^+ e^-$ center-of-mass system (CMS). The $B$ candidates that satisfy $\Mbc > 5.24 \; {\rm GeV}/c^{2}$ and $-0.20 \; {\rm GeV} < \De < 0.15 \; {\rm GeV}$ are retained for further analysis.

The dominant background in the reconstruction of \Brec\ arises from continuum production. Since continuum events tend to be jetlike, in contrast to spherical \BBbar\ decays, continuum background can be distinguished from \BBbar\ signal using event-shape variables, which we combine into a Fisher discriminant \Fsb~\cite{nazi_stuff}. The \BBbar\ training sample is taken from signal MC, while the \qqbar\ training sample is from the off-resonance data sample. The Fisher discriminant is then constructed from the variables described in Ref.~\cite{a1pi_Belle}. The variable providing the strongest discrimination against continuum is the cosine of the angle between the \Brec\ thrust direction (TB) and the thrust of the tag side (TO) $|\cos \theta_{\rm TB, TO}|$. The thrust is defined as the vector that maximizes the sum of the longitudinal momenta of the particles. For a \BBbar\ event, the pair is nearly at rest in the CMS, so the thrust axis of \Brec\ is uncorrelated with the thrust axis of \Btag. In a \qqbar\ event, on the other hand, the decay products align along two nearly back-to-back jets, so the two thrust axes tend to be collinear. Before training, a loose requirement of $|\cos \theta_{\rm TB, TO}| < 0.9$ is imposed that retains 90\% of the signal while rejecting 50\% of the continuum background. The range of the Fisher discriminant $-3 < \Fsb < 2$ encompasses all signal and background events.

Backgrounds from charm ($b \to c$) decays are found to be negligible and are thus not considered, while charmless ($b \to u,d,s$) decays of the $B$ meson may contribute, though rarely in the same region of \Mbc\ and \De\ where signal is present.

As the \Brec\ and \Btag\ are almost at rest in the \Ups\ CMS, the difference in decay time between the two $B$ candidates, $\Delta t$, can be determined approximately from the displacement in $z$ between the final state decay vertices as
\begin{equation}
  \Dt \simeq \frac{(z_{\rm Rec} - z_{\rm Tag})}{\beta \gamma c} \equiv \frac{\Delta z}{\beta \gamma c}.
\end{equation}

The vertex of reconstructed $B$ candidates is determined from the charged daughters, with a further constraint coming from the known IP. The IP profile is smeared in the plane perpendicular to the $z$ axis to account for the finite flight length of the $B$ meson in that plane. To obtain the \Dt\ distribution, we reconstruct the tag side vertex from the tracks not used to reconstruct \Brec~\cite{ResFunc}. Candidate events must satisfy the requirements $|\Dt| < 70 \; {\rm ps}$ and $h_{\rm Rec, Tag} < 500$, where $h_{\rm Rec, Tag}$ is the multitrack vertex goodness-of-fit, calculated in three-dimensional space without using the IP profile constraint~\cite{jpsiks_Belle2}. To avoid the necessity of also modeling the event-dependent observables that describe the \Dt\ resolution in the fit
~\cite{Punzi}, the vertex uncertainty is required to satisfy the loose criteria $\sigma^{\rm Rec, Tag}_z < 200 \; \mu {\rm m}$ for multitrack vertices and $\sigma^{\rm Rec, Tag}_z < 500 \; \mu {\rm m}$ for single-track vertices.

The flavor tagging procedure is described in Ref.~\cite{Tagging}. The tagging information is represented by two parameters, the \Btag\ flavor $q$ and the flavor-tagging quality $r$. The parameter $r$ is continuous and determined on an event-by-event basis with an algorithm trained on MC simulated events, ranging from zero for no flavor discrimination to unity for an unambiguous flavor assignment. To obtain a data-driven replacement for $r$, we divide it into seven regions and determine a probability of mistagging $w$ for each $r$ region using high statistics control samples. Due to a nonzero probability of mistagging $w$, the $CP$ asymmetry in data is thus diluted by a factor $1-2w$ instead of the MC-determined $r$. The measure of the flavor tagging algorithm performance is the total effective tagging efficiency $\epsilon_{\rm eff} = \epsilon_{\rm Tag}(1-2w)^2$, rather than the raw tagging efficiency $\epsilon_{\textrm{Tag}}$, as the statistical significance of the $CP$ parameters is proportional to $(1-2w)\sqrt{\epsilon_{\rm Tag}}$. These are determined from data to be $\epsilon_{\rm eff} = 0.284\pm0.010$ and $\epsilon_{\rm eff} = 0.301\pm0.004$ for the SVD1 and SVD2 data, respectively~\cite{jpsiks_Belle2}.

About 1\% of events have more than one $B$ candidate. For these events, the candidate containing the two highest momentum tracks in the lab frame is selected.

Differences from the previous Belle analysis~\cite{pipi_Belle} include an improved tracking algorithm that was applied to the SVD2 data sample and the inclusion of the event shape \Fsb\ into the fit rather than the optimization of selection criteria for this variable. As the latter strategy results in a large increase of the continuum background level, a reduced fit region in \Mbc\ and \De\ is chosen in order to reduce this background without significant loss of signal events. According to MC simulation, these changes increase the detection efficiency by 19\% over the previous analysis at a cost of continuum levels rising 4.7 times higher in the signal region defined by the previous analysis.

\section{Event Model}
\label{Event Model}
The $CP$ violation parameters are extracted from a seven-dimensional unbinned extended maximum likelihood fit to \Mbc, \De, \Fsb, \Lpm, \Dt\ and $q$ from a data sample divided into seven bins ($l = 0..6$) in the flavor-tag quality $r$ and 2 SVD configurations $s$. Seven categories are considered in the event model: \pippim\ signal, \kppim, \kmpip\ and \kpkm\ peaking backgrounds, continuum, charmless neutral and charged $B$ decays. For most categories, the linear correlations between fit variables are small, so the probability density function (PDF) for each category $j$ is taken as the product of individual PDFs for each variable: ${\cal P}^{l,s}_{j}(\Mbc,\De,\Fsb,\Lp,\Lm,\Dt,q) = {\cal P}^{l,s}(\Mbc) \times {\cal P}^{l,s}(\De) \times {\cal P}^{l,s}(\Fsb) \times {\cal P}^{l,s}(\Lp,\Lm) \times {\cal P}^{l,s}(\Dt,q)$ in each $l,s$ bin, unless stated otherwise.

\subsection{Peaking models}
The four peaking shapes, including the signal, are determined from reconstructed MC events. The PDFs for \Mbc\ and \De\ are taken to be the sum of three Gaussian functions, where the two tail Gaussians are parametrized relative to the core, which incorporates calibration factors that correct for the difference between data and MC simulation. These factors calibrate the mean and width of the core Gaussian component. The PDF for \Fsb\ is taken to be the sum of three Gaussians in each flavor-tag bin $l$, where the shape parameters are identical for all peaking channels. Calibration factors that correct for the shape differences between data and MC are incorporated into the core mean and width. These factors for \Mbc\ are determined directly in the fit, while for \De\ and \Fsb, these factors are determined from a large-statistics control sample of \dpi\ decays. The \Lpm\ shape is modeled with a two-dimensional histogram that has been corrected for the difference between data and MC in PID as determined from an independent study with inclusive $D^{*+} \to D^{0} [\Km \pip] \pip_{\rm slow}$ decays. The PDF of \Dt\ and $q$ for \pippim is given by
\begin{eqnarray}
  {\cal P}^{l,s}_{\pip\pim}(\Dt, q) &\equiv& \frac{e^{-|\Dt|/\taub}}{4\taub} \biggl\{1-q\Dw^{l,s}+q(1-2w^{l,s})\times \nonumber \\
  & & \biggl[\Acp\cos \Dmd \Dt + \Scp \sin \Dmd \Dt\biggr]\biggr\} \otimes R^{s}_{\BzBzb}(\Dt),
\end{eqnarray}
which accounts for $CP$ dilution from the probability of incorrect flavor tagging $w^{l,s}$ and the wrong tag difference $\Dw^{l,s}$ between \Bz\ and \Bzb, both of which are determined from flavor-specific control samples using the method described in Ref~\cite{Tagging}. The physics parameters \taub\ and \Dmd\ are fixed to their respective current world averages~\cite{PDG}. This PDF is convolved with the \Dt\ resolution function for neutral $B$ particles $R^{s}_{\BzBzb}$, as in Ref.~\cite{jpsiks_Belle2}. We consider the \Dt,$q$ distributions for the flavor-specific \kppim\ and \kmpip\ peaking backgrounds separately with
\begin{equation}
  {\cal P}^{l,s}_{\Kpm\pimp}(\Dt, q) \equiv \frac{e^{-|\Dt|/\taub}}{4\taub} \biggl\{1-q\Dw^{l,s} \mp q(1-2w^{l,s}) \cos \Dmd \Dt\biggr\} \otimes R^{s}_{\BzBzb}(\Dt).
\end{equation}
For the \kpkm\ peaking background, the \Dt,$q$ PDF is taken to be the same as that for \pippim\ signal, but as \kpkm\ has not yet been observed, the $CP$ parameters are set to zero. To account for the outlier \Dt\ events not described by the \Dt\ resolution function, a broad Gaussian PDF is introduced for every category,
\begin{equation}
  {\cal P}^{l,s}_{\rm Out}(\Dt, q) \equiv \frac{1}{2}G(\Dt; 0, \sigma^{s}_{\rm Out}).
\end{equation}

\subsection{Continuum model}
The parametrization of the continuum model is based on the off-resonance data; however, all the shape parameters of \Mbc, \De, \Fsb\ and \Lpm\ are floated in the fit. As continuum is the dominant component, extra care is taken to ensure that this background shape is understood as precisely as possible, incorporating correlations above 2\%. The PDF for \Mbc\ is an empirical ARGUS function~\cite{ARGUS}, while \De\ is modeled by a linear fit in each flavor-tag bin with a slope parametrized by $p^{l,s}_{0}$ and $p^{s}_{1}$, depending linearly on \Fsb,
\begin{equation}
  {\cal P}^{l,s}_{\qqbar}(\De|\Fsb) = 1 + (p^{l,s}_{0} + p^{s}_{1}\Fsb)\De.
\end{equation}
The \Fsb\ shape is observed to shift depending on the PID region, so the PDF is a sum of two Gaussian functions in two PID regions, $\Lpm \leq 0.5$ and (\Lp\ or $\Lm) > 0.5$. A small correlation between the \Lpm\ shape and flavor-tag $q$ is also observed due to the $s\bar s$ component of continuum. As an example, consider the case where two jets are produced in which one contains a \Kp\ and the other contains a \Km. If a \Brec\ candidate is successfully reconstructed with the \Kp, it inhabits the flavor-specific $\Kp\pim$ sector of \Lpm. Then the accompanying \Km\ could then be used as part of the flavor-tagging routine, which leads to a preferred flavor tag of \Bzb. This enhances the \Lpm\ distribution in the $\Kp\pim$ region and depletes it in the $\Km\pip$ region for $q=-1$. To account for this effect, we model \Lpm\ with an effective asymmetry $A^{l,s}_{\qqbar}$ that modifies the two-dimensional PID histogram model $H^{l,s}(\Lp,\Lm)$, in each $l,s$ bin depending on the flavor tag,
\begin{equation}
  {\cal P}^{l,s}_{\qqbar}(\Lpm,q) = \frac{1+qA^{l,s}_{\qqbar}(\Lp,\Lm)}{2}H^{l,s}(\Lp,\Lm),
\end{equation}
where
\begin{eqnarray}
  A^{l,s}_{\qqbar}(\Lp,\Lm) &=& +a_{0}^{l,s}|\Lm-\Lp|^{a^{s}_{1}} \hspace{20pt} \textrm{if } \Lm-\Lp \geq 0 \nonumber\\
  &=& -a_{0}^{l,s}|\Lm-\Lp|^{a^{s}_{1}} \hspace{20pt} \textrm{if } \Lm-\Lp < 0
  ,
  \label{manta_ray}
\end{eqnarray}
which we hereafter refer to as the ``manta ray'' function. The \Dt\ model,
\begin{equation}
  P^{l,s}_{\qqbar}(\Dt) \equiv \biggl[(1 - f_{\delta}) \frac{e^{-|\Dt|/\tau_{\qqbar}}}{2\tau_{\qqbar}} + f_{\delta} \; \delta( \Dt - \mu^{s}_{\delta})\biggr] \otimes R^{s}_{\qqbar}(\Dt),
\end{equation}
contains a lifetime and prompt component to account for the charmed and charmless contributions, respectively. It is convolved with a sum of two Gaussians,
\begin{equation}
  R^{s}_{\qqbar}(\Dt) \equiv (1-f^{s}_{\rm tail})G(\Dt; \mu^{s}_{\rm mean}, S^{s}_{\rm main}\sigma) + f^{s}_{\rm tail}G(\Dt; \mu^{s}_{\rm mean}, S^{s}_{\rm main}S^{s}_{\rm tail}),
\end{equation}
which uses the event-dependent \Dt\ error constructed from the estimated vertex resolution $\sigma \equiv (\sqrt{\sigma^{2}_{\rm Rec}+\sigma^{2}_{\rm Tag}})/\beta \gamma c$ as a scale factor of the width parameters $S^{s}_{\rm main}$ and $S^{s}_{\rm tail}$.

\subsection{\mbox{\boldmath${\BBbar}$} model}
The charmless $B$ background shape is determined from a large sample of MC events based on $b \to u,d,s$ transitions that is further subdivided into neutral and charged $B$ samples. A sizeable correlation of 18\% is found between \Mbc\ and \De\ and is taken into account with a two-dimensional histogram. The PDF for \Fsb\ is taken to be the sum of three Gaussians in each flavor-tag bin $l$, similar to the peaking model. Here, we are able to fix the shape parameters from the peaking model except for the core mean and width. A similar correlation between the flavor tag and \Lpm, similar to that in continuum, is also observed. Due to \BzBzb\ mixing in the neutral $B$ background, this effect is correlated with $\Dt$ and $q$. For the neutral $B$ background, the PDF is given by
\begin{eqnarray}
  {\cal P}^{l,s}_{\BzBzb}(\Lpm,\Dt, q) &=& H^{l,s}(\Lp,\Lm) \times \nonumber \\
  && \frac{e^{-|\Dt|/\tau_{\BzBzb}}}{4\tau_{\BzBzb}} \biggl\{1 + q A^{l,s}_{\BzBzb}(\Lp,\Lm) \cos \Dmd \Dt\biggr\} \otimes R^{s}_{\BzBzb}(\Dt), \nonumber \\
\end{eqnarray}
and the charged $B$ background PDF is given by
\begin{equation}
  {\cal P}^{l,s}_{\BpBm}(\Lpm,\Dt, q) = \frac{1+qA^{l,s}_{\BpBm}(\Lp,\Lm)}{2}H^{l,s}(\Lp,\Lm) \frac{e^{-|\Dt|/\tau_{\BpBm}}}{2\tau_{\BpBm}} \otimes R^{s}_{\BpBm}(\Dt),
\end{equation}
where $A^{l,s}_{\BBbar}$
 are manta ray functions
for each \BBbar\ category and $R_{\BpBm}$ is the \Dt\ resolution function for charged $B$ events. As reconstructed background $B$ candidates may borrow a track from the tag side, the average \Dt\ lifetime tends to be smaller and is taken into account with the effective lifetime, $\tau_{\BBbar}$.

\subsection{Full model}
The total likelihood for $559797$ \hphm\ candidates in the fit region is
\begin{equation}
  {\cal L} \equiv \prod_{l,s} \frac{e^{-\sum_{j}N^{s}_{j}\sum_{l,s}f^{l,s}_{j}}}{N_{l,s}!} \prod^{N_{l,s}}_{i=1} \sum_{j}N^{s}_{j}f^{l,s}_{j}{\cal P}^{l,s}_{j}(\Mbc^{i}\De^{i},\Fsb^{i},{\cal L}_{K/\pi}^{+ \; i},{\cal L}_{K/\pi}^{- \; i},\Dt^{i},q^{i}),
  \label{eq_likelihood}
\end{equation}
which iterates over $i$ events, $j$ categories, $l$ flavor-tag bins and $s$ detector configurations. The fraction of events in each $l,s$ bin, for category $j$, is denoted by $f^{l,s}_{j}$. The fraction of signal events in each $l,s$ bin, $f^{l,s}_{\rm Sig}$, is calibrated with the \dpi\ control sample. Free parameters of the fit include the \pippim\ and \kpkm\ yields, $N^{s}_{\qqbar}$ and $N^{s}_{\BzBzb}$. The individual \kppim\ and \kmpip\ yields are parametrized in terms of their combined yield $N_{K\pi}$ and the $CP$ violating parameter $\Acp^{K\pi}$, which are both free in the fit: $N_{K^{\pm}\pi^{\mp}} = N_{K\pi}(1 \mp \Acp^{K\pi})/2$. The remaining $N^{s}_{\BpBm}$ yields are fixed to $N^{\rm SVD1}_{\BpBm} = (0.269\pm0.010)N^{\rm SVD1}_{\BzBzb}$ and $N^{\rm SVD2}_{\BpBm} = (0.268\pm0.004)N^{\rm SVD2}_{\BzBzb}$ as determined from MC simulation. In addition, all shape parameters of the continuum model with the exception of the \Dt\ parameters are allowed to vary in the fit. In total, there are 116 free parameters in the fit: 10 for the peaking models, 104 for the continuum shape and 2 for the \BBbar\ background.

To determine the component yields and $CP$ violation parameters, in contrast to the previous Belle analysis~\cite{pipi_Belle}, we fit all variables simultaneously. The previous analysis applied a two-step procedure where the event-dependent component probabilities were calculated from a fit without \Dt\ and $q$. These were then used as input in a fit to \Dt\ and $q$ to set the fractions of each component to determine the $CP$ parameters. Our procedure has the added benefit of further discrimination against continuum with the \Dt\ variable and makes the treatment of systematic uncertainties more straightforward, at a cost of analysis complexity and longer computational time. A pseudoexperiment study indicates a 10\% improvement in statistical uncertainty of the $CP$ parameters over the previous analysis method.

\section{Results}
\label{Fit Result}
\begin{figure}
  \centering
  \includegraphics[height=180pt,width=!]{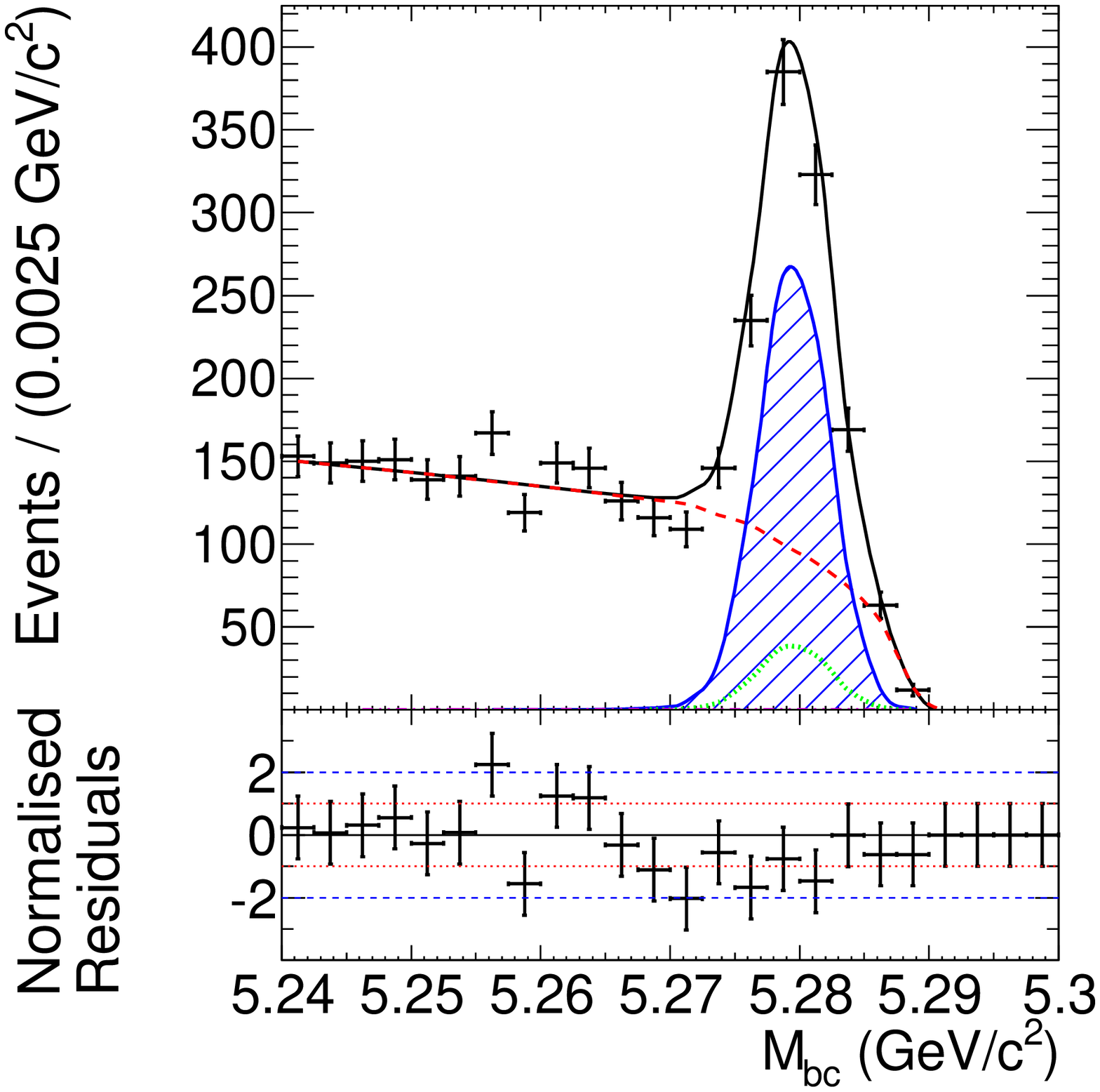}
  \includegraphics[height=180pt,width=!]{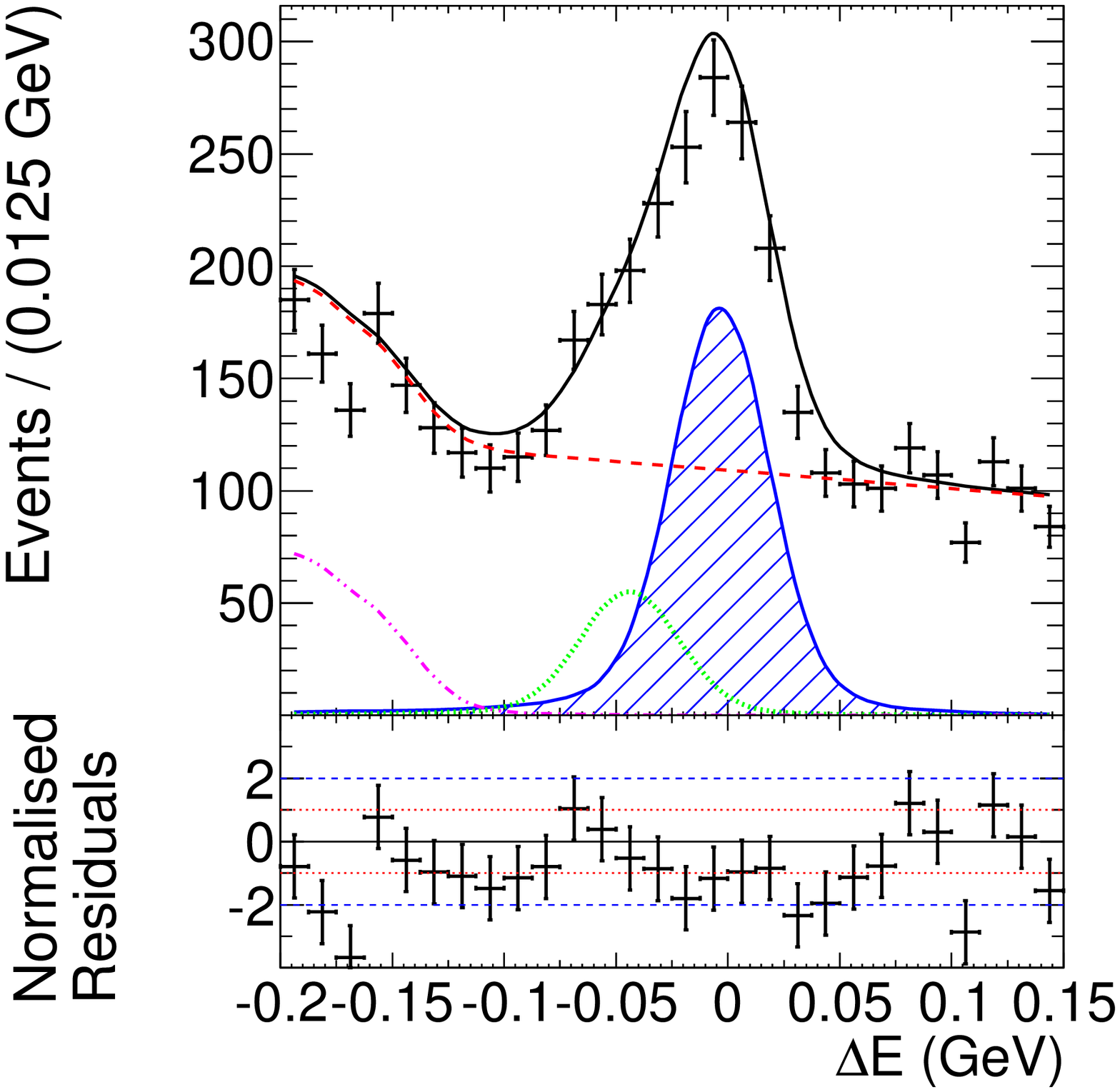}
  \put(-227,155){(a)}
  \put(-40,155){(b)}

  \includegraphics[height=180pt,width=!]{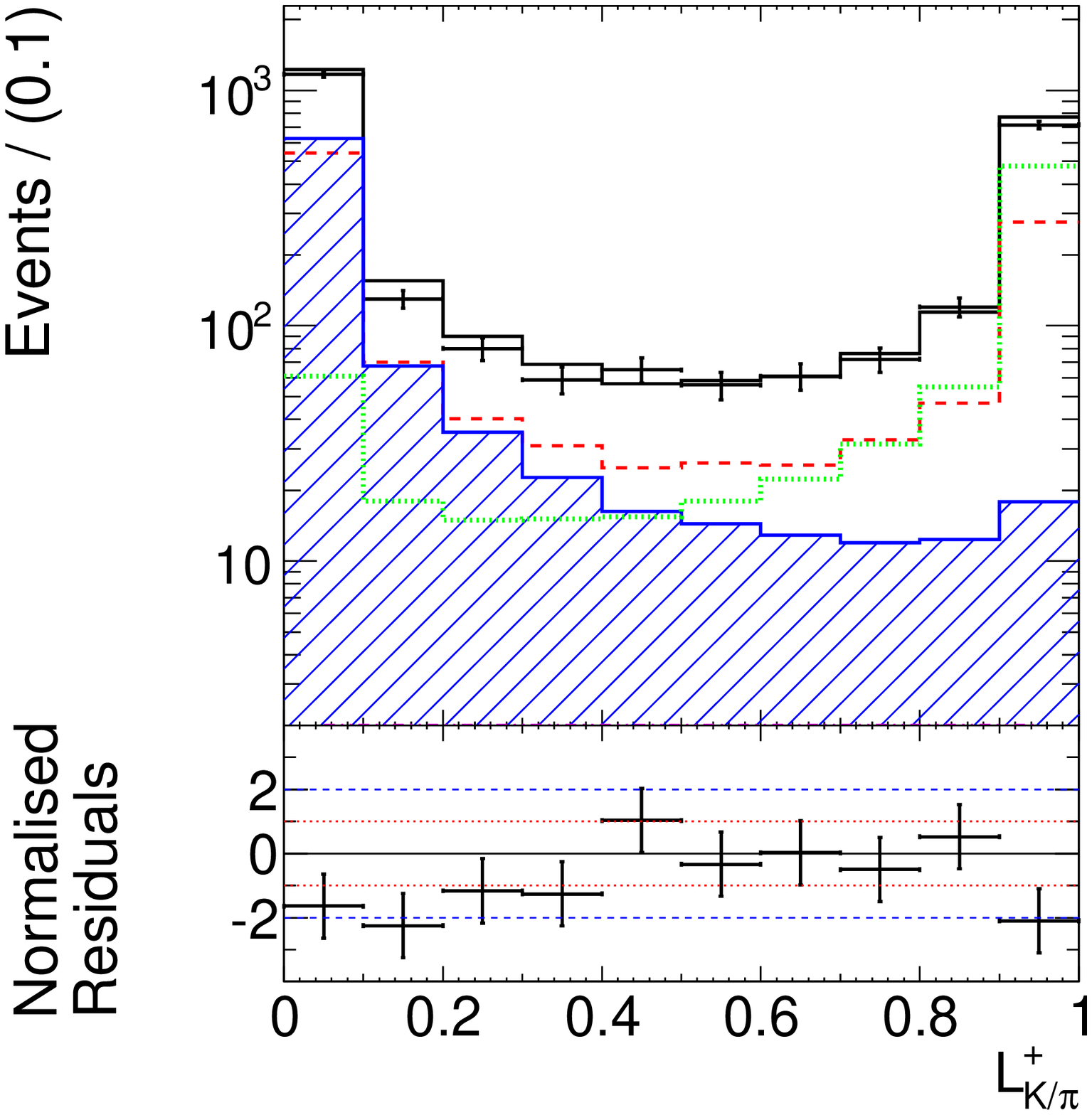}
  \includegraphics[height=180pt,width=!]{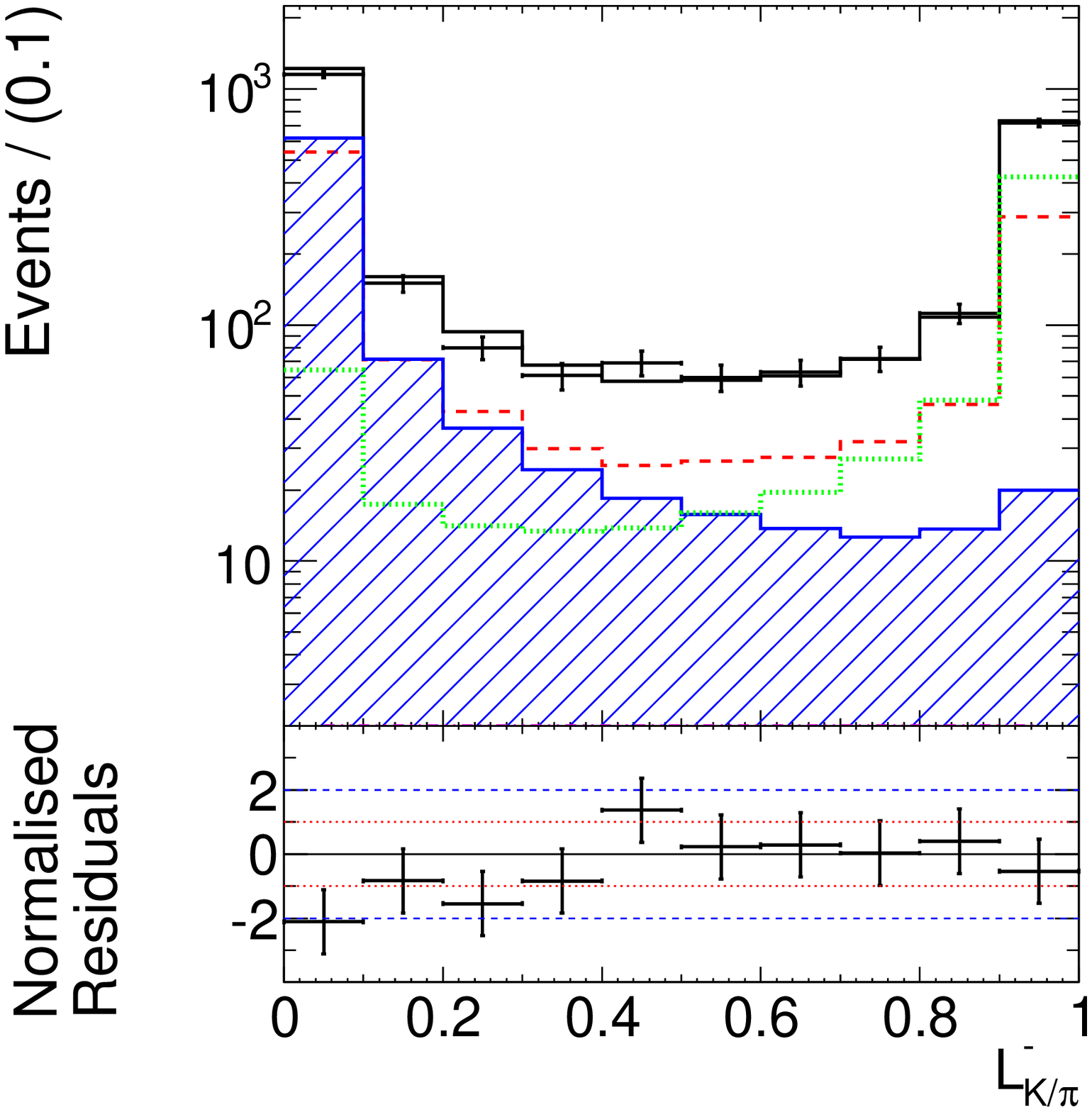}
  \put(-227,155){(c)}
  \put(-40,155){(d)}

  \includegraphics[height=180pt,width=!]{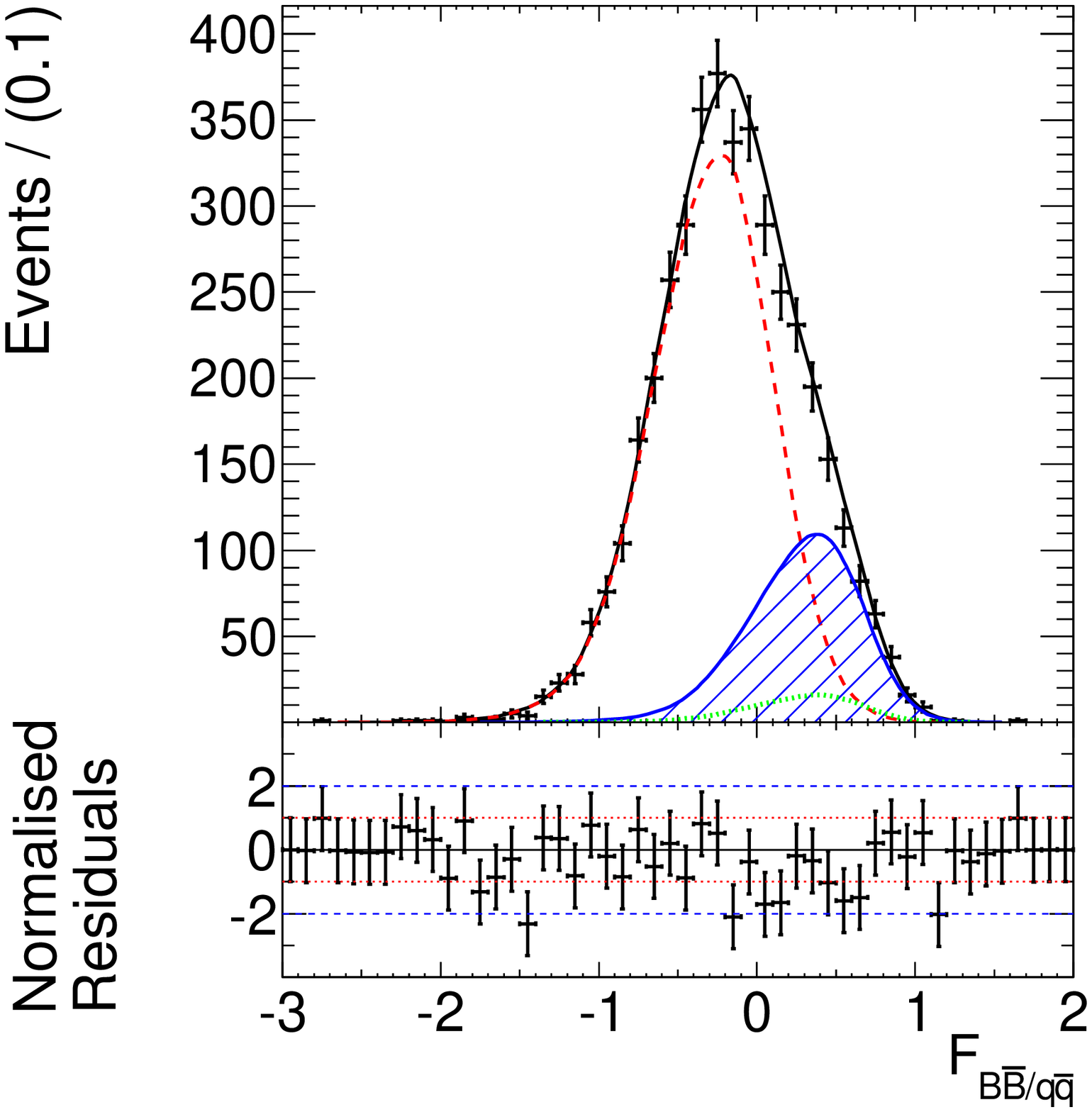}
  \put(-40,155){(e)}

  \caption{(color online) Projections of the fit to the data enhanced in the \pippim\ signal region. Points with error bars represent the data and the solid black curves or histograms represent the fit results. The signal enhancements, $\Mbc > 5.27 \textrm{ GeV}/c^2$, $|\De| < 0.04 \textrm{ GeV}$, $\Fsb > 0$, $\Lpm < 0.4$ and $r > 0.5$, except for the enhancement of the dimension being plotted are applied to each projection. (a), (b), (c), (d) and (e) show the \Mbc, \De, \Lp, \Lm\ and \Fsb\ projections, respectively. Blue hatched curves show the \pippim\ signal component, green dotted curves show the \kpi\ peaking background component, dashed red curves indicate the total background, and purple dash-dotted curves show the \BBbar\ background component.}
  \label{fig_data_pipi1}
\end{figure}

\begin{figure}
  \centering
  \includegraphics[height=200pt,width=!]{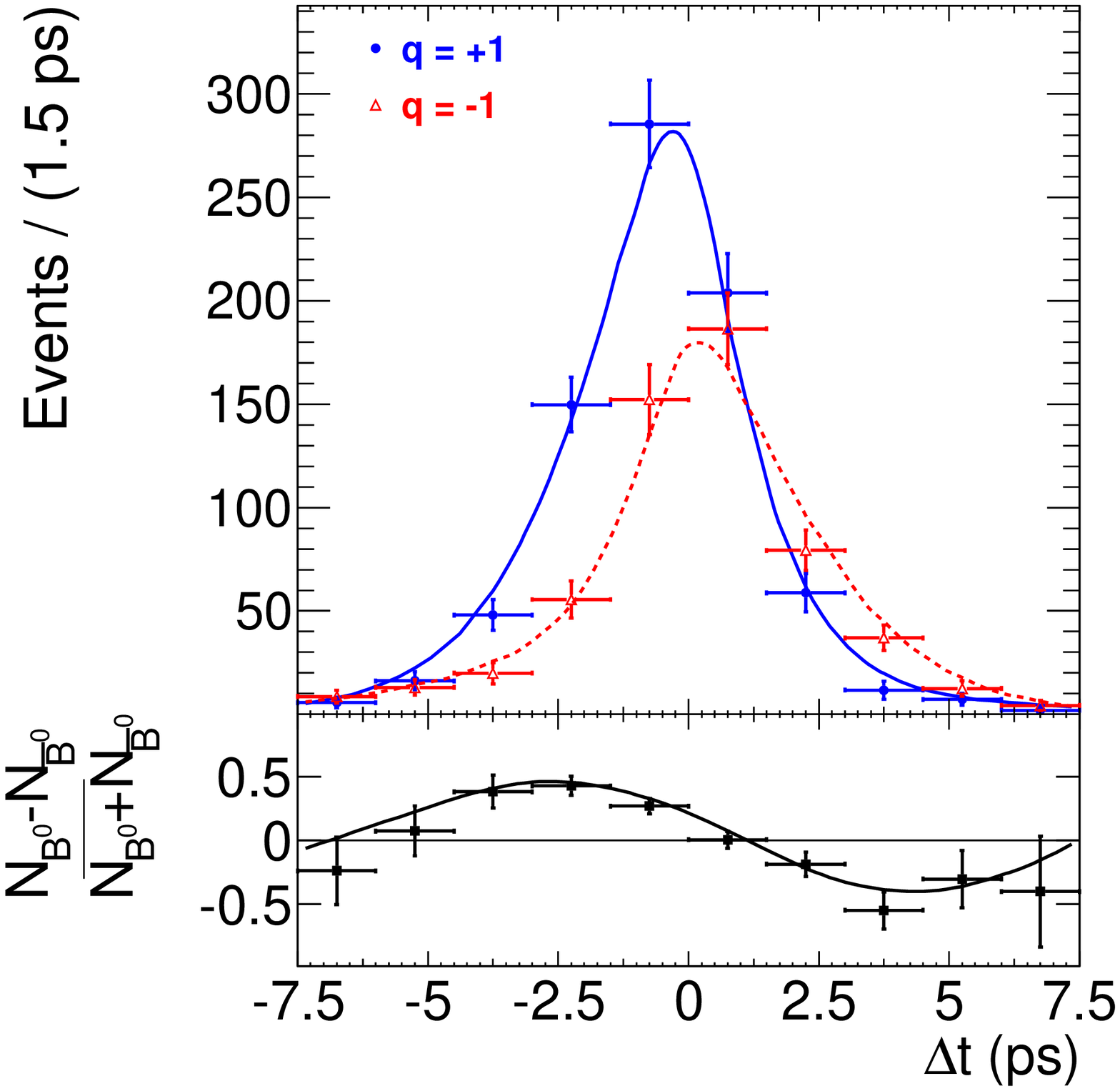}
  \put(-42,173){(a)}
  \put(-42,53){(b)}

  \caption{(color online) Background subtracted time-dependent fit results for \pippim. (a) shows the \Dt\ distribution for each \Btag\ flavor $q$. The solid blue and dashed red curves represent the \Dt\ distributions for \Bz\ and \Bzb\ tags, respectively. (b) shows the asymmetry of the plot above them, $(N_{\Bz} - N_{\Bzb})/(N_{\Bz} + N_{\Bzb})$, where $N_{\Bz}$ ($N_{\Bzb}$) is the measured signal yield of \Bz\ (\Bzb) events in each bin of \Dt.}
  \label{fig_data_pipi2}
\end{figure}

From the fit to the data, the following $CP$ violation parameters are obtained:
\begin{eqnarray}
  \Acp(\pippim) &=& +0.33 \pm 0.06 \textrm{ (stat)} \pm 0.03 \textrm{ (syst)},\nonumber\\
  \Scp(\pippim) &=& -0.64 \pm 0.08 \textrm{ (stat)} \pm 0.03 \textrm{ (syst)},
\end{eqnarray}
where the first uncertainty is statistical and the second is the systematic error (Sec.~\ref{Systematic Uncertainties}). Signal-enhanced fit projections are shown in Figs.~\ref{fig_data_pipi1} and \ref{fig_data_pipi2}. The effects of neglecting the correlation between \Mbc\ and \De\ in the peaking models can be seen there as the slight overestimation of signal; however, pseudoexperiments show that this choice does not bias the $CP$ violation parameters. These results are the world's most precise measurements of time-dependent $CP$ violation parameters in \pippim. The statistical correlation coefficients between the $CP$ violation parameters is $+0.10$. The peaking event yields including signal are $N(\pippim) = 2964 \pm 88$, $N(\kppim) = 9205 \pm 124$ and $N(\kpkm) = 23 \pm 35$, where the uncertainties are statistical only. From the yields obtained in the fit, the relative contributions of each component are found to be $0.5\%$ for \pippim, $1.6\%$ for \kppim, $97.7\%$ for continuum and $0.2\%$ for \BBbar\ background. For the $CP$ violating parameter $\Acp^{K\pi}$, we obtain a value of $-0.061 \pm 0.014$, which is consistent with the latest Belle measurement~\cite{hphm_Belle}.

Our results confirm $CP$ violation in this channel as reported in previous measurements and other experiments~\cite{pipi_Belle,pipi_BaBar,pipi_LHCb}, and the value for \Acp\ is in marginal agreement with the previous Belle measurement. As a test of the accuracy of the result, we perform a fit on the data set containing the first $535 \times 10^6$ \BBbar\ pairs, which corresponds to the integrated luminosity used in the previous analysis. We obtain $\Acp = +0.47 \pm 0.07$ which is in good agreement with the value shown in Table~\ref{tab_hh_prev}, considering the new tracking algorithm and the 19\% increase in detection efficiency due to improved analysis strategy. In a separate fit to only the new data sample containing $237 \times 10^6$ \BBbar\ pairs, we obtain $\Acp = +0.06 \pm 0.10$. Using a pseudoexperiment technique based on the fit result, we estimate the probability of a statistical fluctuation in the new data set causing the observed shift in central value of \Acp\ from our measurement with the first $535 \times 10^6$ \BBbar\ pairs to be 0.5\%.

To test the validity of the \Dt\ resolution description, we perform a separate fit with a floating \Bz\ lifetime; the result for $\tau_{\Bz}$ is consistent with the current world average~\cite{PDG} within $2\sigma$. As a further check of the \Dt\ resolution function and the parameters describing the probability of mistagging, we fit for the $CP$ parameters of our control sample \dpi; the results are consistent with the expected null asymmetry. Finally, we determine a possible fit bias from a MC study in which the peaking channels and \BBbar\ backgrounds are obtained from GEANT-simulated events, and the continuum background is generated from our model of off-resonance data. The statistical errors observed in this study agree with those obtained from our fit to the data.

\begin{figure}
  \centering
  \includegraphics[height=160pt,width=!]{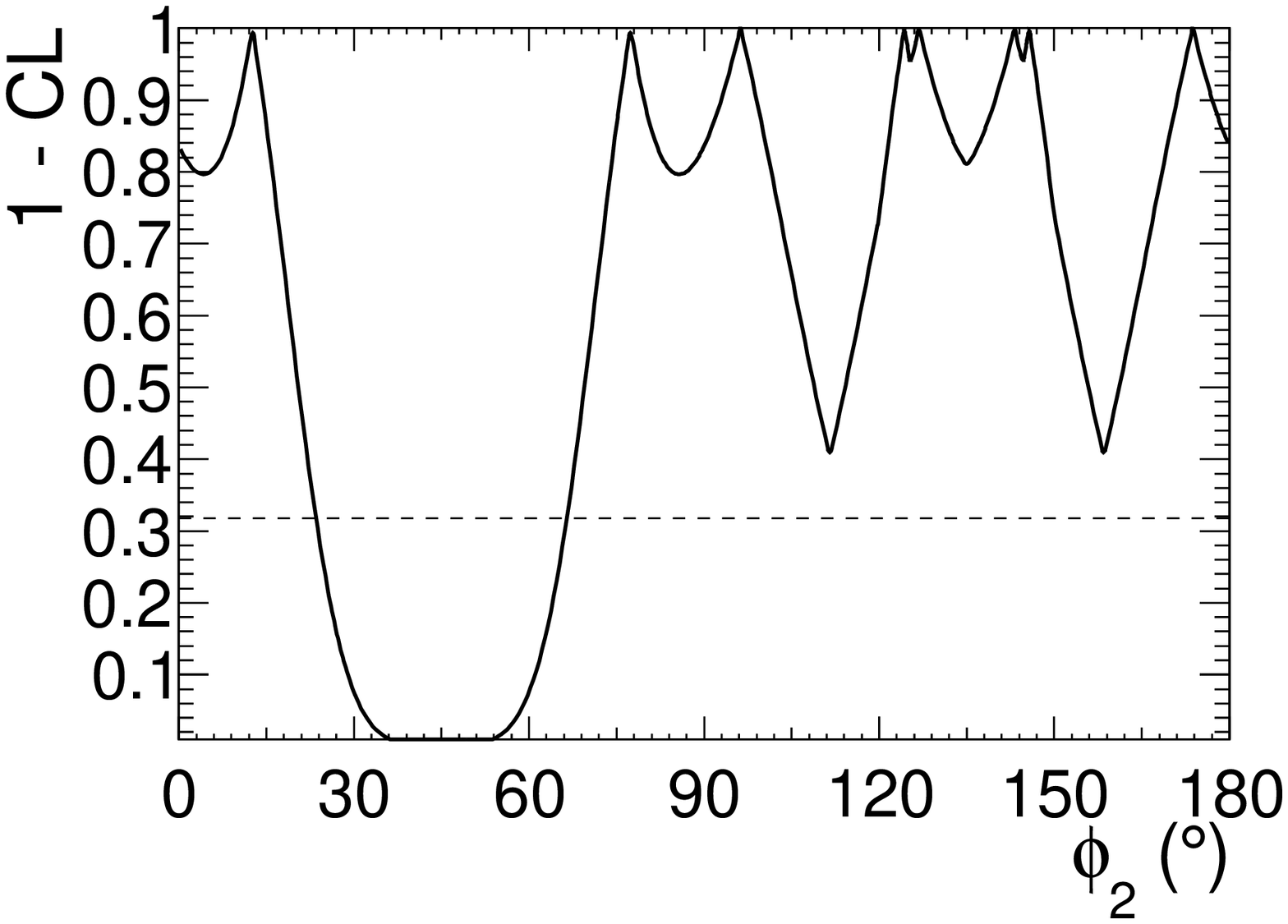}
  \includegraphics[height=160pt,width=!]{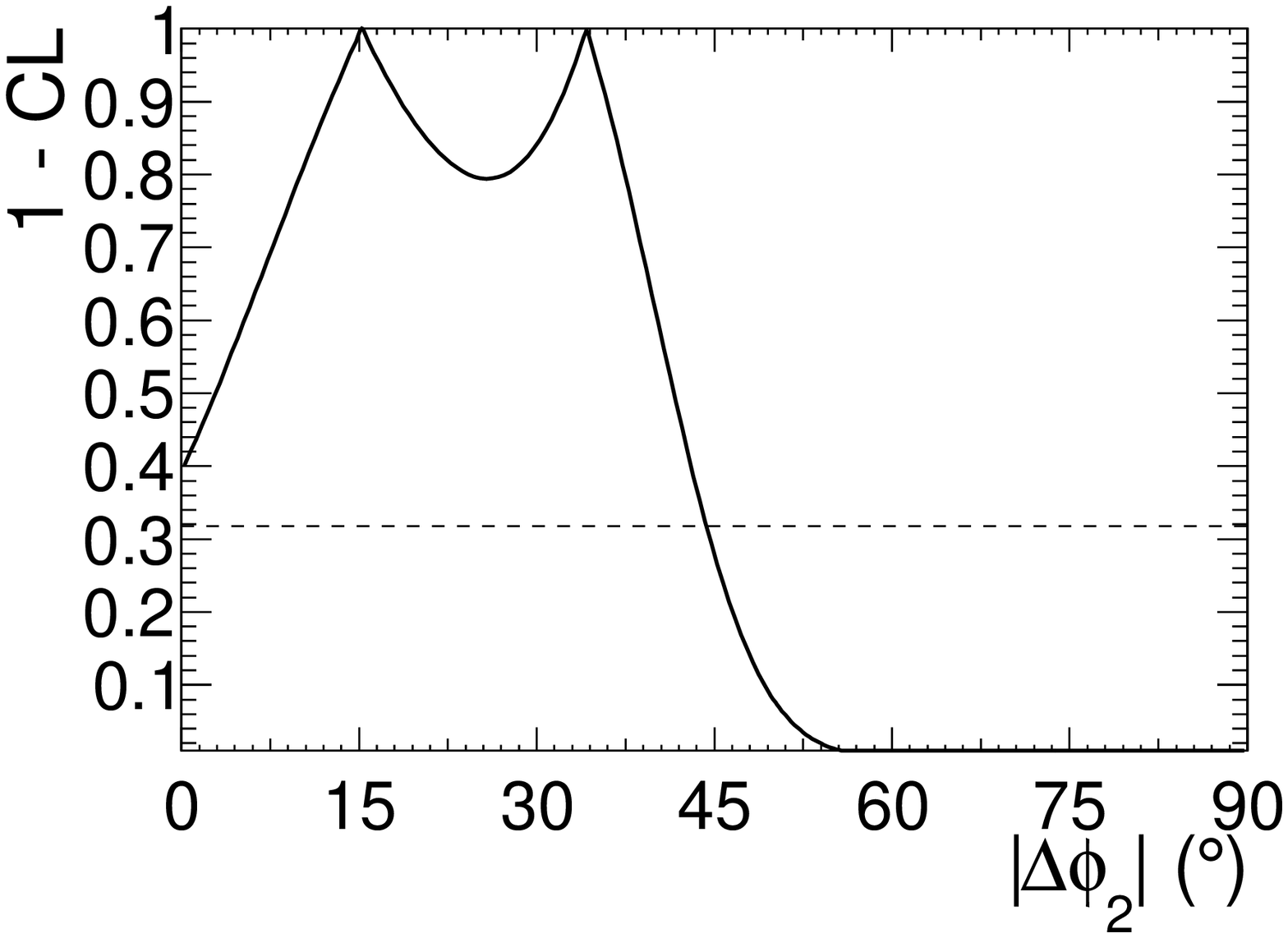}
  \put(-268,40){(a)}
  \put(-45,40){(b)}

  \caption{Difference 1-CL, plotted for a range of \phitwo\ (a) and (b) $|\Delta \phitwo|$ values as shown by the solid curve. The dashed lines indicate the $1\sigma$ exclusion level.}
  \label{fig_CL}
\end{figure}

Using Eq.~(\ref{eq_iso}) and input from other Belle publications~\cite{hphm_Belle,pi0pi0_Belle}, an isospin analysis is performed to constrain the angle \phitwo. A goodness-of-fit $\chi^{2}$ is constructed for the five amplitudes shown in Fig.~\ref{fig_iso}, accounting for the correlations between our measured physics observables used as input. The $\chi^{2}$ is then converted into a $p$ value (CL) as shown in Fig.~\ref{fig_CL}. The region $23.8^{\circ} < \phitwo < 66.8^{\circ}$ is disfavored and the constraint on the shift in \phitwo\ caused by the penguin contribution is $|\Delta\phitwo| < 44.8^{\circ}$ at the $1\sigma$ level, including systematic uncertainties.

\section{Systematic Uncertainties}
\label{Systematic Uncertainties}
Systematic errors from various sources are considered and estimated with independent internal studies and cross-checks. These are summarized in Table~\ref{tab_syst}. Uncertainties affecting the vertex reconstruction include the IP profile, charged track selection based on track helix errors, helix parameter corrections, \Dt\ and vertex goodness-of-fit selection, \Dz\ bias and SVD misalignment. The fit model uncertainties including the fixed physics parameters \taub\ and \Dmd, parameters describing the difference between data and MC simulation, \Dt\ resolution function parameters, as well as the flavor-tagging performance parameters $w$ and \Dw, are varied by $\pm 1 \sigma$. The parametric and nonparametric shapes describing the background are varied within their uncertainties. For nonparameteric shapes ({\it i.e.}, histograms), we vary the contents of the histogram bins by $\pm 1\sigma$. The fit bias is determined from the difference between the generated and fitted physics parameters using pseudoexperiments. Finally, a large number of MC pseudoexperiments are generated and an ensemble test is performed to obtain possible systematic biases from interference on the tag side arising between the CKM-favored $b \bar d \to (c \bar u d) \bar d$ and doubly CKM-suppressed $\bar b d \to (\bar u c \bar d) d$ amplitudes in the final states used for flavor tagging~\cite{tsi}.

\begin{table}
  \footnotesize
  \centering
  \caption{Systematic uncertainties of the measured physics parameters.}
  \begin{tabular}
    {@{\hspace{0.5cm}}c@{\hspace{0.25cm}}  @{\hspace{0.25cm}}c@{\hspace{0.5cm}}  @{\hspace{0.25cm}}c@{\hspace{0.5cm}}}
    \hline \hline
    Category & $\delta\Acp(\pip\pim)$ $(10^{-2})$ & $\delta\Scp(\pip\pim)$ $(10^{-2})$\\
    \hline
    IP profile & 0.13 & 1.19\\
    \Btag\ track selection & 0.30 & 0.33\\
    Track helix errors & 0.00 & 0.01\\
    \Dt\ selection & 0.01 & 0.03\\
    Vertex quality selection & 0.37 & 0.23\\
    $\Delta z$ bias & 0.50 & 0.40\\
    Misalignment & 0.40 & 0.20\\
    \taub\ and \Dmd & 0.12 & 0.09\\
    Data/MC shape & 0.15 & 0.19\\
    \Dt\ resolution function & 0.83 & 2.02\\
    Flavor tagging & 0.40 & 0.31\\
    Background Parametric shape & 0.15 & 0.28\\
    Background Nonparametric shape & 0.37 & 0.57\\
    Fit bias & 0.54 & 0.86\\
    Tag-side interference & 3.18 & 0.17\\\hline
    Total & 3.48 & 2.68\\
    \hline \hline
  \end{tabular}
  \label{tab_syst}
\end{table}

\section{Conclusion}
\label{Conclusion}
We report an improved measurement of the $CP$ violation parameters in \pippim\ decays, confirming $CP$ violation in this channel as reported in previous measurements and other experiments~\cite{pipi_Belle,pipi_BaBar,pipi_LHCb}. These results are based on the full Belle data sample after reprocessing with a new tracking algorithm and with an optimized analysis performed with a single simultaneous fit, and they supersede those of the previous Belle analysis~\cite{pipi_Belle}. They are now the world's most precise measurement of time-dependent $CP$ violation parameters in \pippim, disfavoring the range $23.8^{\circ} < \phitwo < 66.8^{\circ}$, at the $1\sigma$ level.

\section*{ACKNOWLEDGMENTS}

We thank the KEKB group for the excellent operation of the
accelerator; the KEK cryogenics group for the efficient
operation of the solenoid; and the KEK computer group,
the National Institute of Informatics, and the 
PNNL/EMSL computing group for valuable computing
and SINET4 network support.  We acknowledge support from
the Ministry of Education, Culture, Sports, Science, and
Technology (MEXT) of Japan, the Japan Society for the 
Promotion of Science (JSPS), and the Tau-Lepton Physics 
Research Center of Nagoya University; 
the Australian Research Council and the Australian 
Department of Industry, Innovation, Science and Research;
Austrian Science Fund under Grant No. P 22742-N16;
the National Natural Science Foundation of China under
Contract No.~10575109, No. 10775142, No. 10875115 and No. 10825524; 
the Ministry of Education, Youth and Sports of the Czech 
Republic under Contract No.~MSM0021620859;
the Carl Zeiss Foundation, the Deutsche Forschungsgemeinschaft
and the VolkswagenStiftung;
the Department of Science and Technology of India; 
the Istituto Nazionale di Fisica Nucleare of Italy; 
the BK21 and WCU program of the Ministry of Education, Science and
Technology; the National Research Foundation of Korea Grants No.\ 
2010-0021174, No. 2011-0029457, No. 2012-0008143, No. 2012R1A1A2008330;
the BRL program under NRF Grant No. KRF-2011-0020333;
the GSDC of the Korea Institute of Science and Technology Information;
the Polish Ministry of Science and Higher Education and 
the National Science Center;
the Ministry of Education and Science of the Russian
Federation and the Russian Federal Agency for Atomic Energy;
the Slovenian Research Agency;
the Basque Foundation for Science (IKERBASQUE) and the UPV/EHU under 
program UFI 11/55;
the Swiss National Science Foundation; the National Science Council
and the Ministry of Education of Taiwan; and the U.S.\
Department of Energy and the National Science Foundation.
This work is supported by a Grant-in-Aid from MEXT for 
Science Research in a Priority Area (``New Development of 
Flavor Physics'') and from JSPS for Creative Scientific 
Research (``Evolution of Tau-lepton Physics'').

\end{document}